\documentclass[prl,aps,twocolumn,showpacs]{revtex4}
\usepackage{graphicx}
\usepackage{dcolumn}
\usepackage{amsmath}
\begin{document}
\title{Temperature dependent magnetic properties of FePt: effective spin
Hamiltonian model}

\author{O.\ N.\ Mryasov, U.\ Nowak, K.\ Y.\  Guslienko, and R.\ W.\ Chantrell}

\affiliation{Seagate Research, 1251 Waterfront Place, Pittsburgh,
PA,
  15222, USA}


\begin{abstract}
A model of magnetic interactions in the ordered ferromagnetic FePt
is proposed on the basis of first-principles calculations of
non-collinear magnetic configurations and shown to be capable of
explaining recent measurements of magnetic anisotropy energy (MAE).
The site (Fe,Pt) resolved contributions to the MAE have been
distinguished with small  Fe easy-plane and large Pt easy-axis
terms. This model has been tested against available experimental
data on the  temperature dependence of MAE showing scaling of
uniaxial MAE (K$_{1}$(T)) with magnetization (M(T)) $K_{1}(T) \sim
M(T)^{\gamma}$ characterized by the unusual  exponent of $\gamma=
2.1$. It is shown that this unusual behavior of the FePt can be
quantitatively explained within the proposed model and originates
from an effective anisotropic exchange mediated by the induced Pt
moment. The latter is expected to be a common feature of 3d-5d(4d)
alloys having 5d/4d elements with large spin-orbit coupling and
exchange enhanced Stoner susceptibility.


\end{abstract}


 \maketitle

\section{Introduction}
Since the phenomenon of super-paramagnetism has been identified as
one of the major limits for the conventional magnetic recording
\cite{weller_para}, significant research effort has been invested in
the development of materials with large magnetic anisotropy energy
(MAE). Recent progress in the fabrication and characterization of
granular and nano-particulate FePt films \cite{weller_fept} puts
even more emphasis on the understanding of the giant MAE of FePt and
its temperature dependence. The latter property appears to be of
critical importance for the development of future high density
magnetic recording systems in particular for heat-assisted magnetic
recording \cite{weller_para}.

A systematic understanding of the temperature dependence of the MAE in
itinerant magnets remains a challenge and one of the long standing
problems in the theory of magnetism. The proposed model deals with
mixed localized and itinerant magnetic moments and thus bears general
importance as large anisotropy is achieved by combining strongly
magnetic elements with non-magnetic ones, where the latter have large
spin orbit coupling.

The chemically ordered $L1_{0}$ phase of FePt has large uniaxial MAE
with the first order anisotropy constant $K_{1} \approx 10^{8}$
erg/cc \cite{weller_para} based on the simple angular variation of
MAE $E^{anis} \sim K_{1} \sin^2{\theta}$.  In the $L1_{0}$ phase the
cubic symmetry is broken due to the stacking of alternate planes of
the 3d element (Fe) and the 5d element (Pt) along the [001]
direction. It is well established that in this naturally layered
ferromagnet the large MAE is mainly due to the contribution from the
5d element having large spin-orbit (s-o) coupling while the 3d
element provides the exchange splitting of the 5d sub-lattice
\cite{oppenier,igor,mae_our}.

Theoretical framework for the temperature dependence of the MAE
was mainly developed in the 50-60's. These efforts led to the
formulation of a general quantum statistical perturbation theory
(QSPT) summarized by Callen and Callen \cite{cal_rev}. This theory
provides a general approach for calculating the magnetic
anisotropy free energy for the effective spin Hamiltonian $H =
H^{iso} + H^{anis}$ with a large isotropic $H^{iso}$ and a
relatively small anisotropic part $H^{anis}$. The anisotropic part
is taken in the form $ H^{anis} = \sum_{i} k_{2} \it{L}_{2}({\bf
S}_{i})$, where $\it{L}_{2}({\bf S}_{i})$ is a normalized
polynomial of $2^{\mathrm{nd}}$ order in the case of uniaxial
symmetry with ${\bf S}_{i}$ denoting a unit vector (in the
classical case) at the atomic site $i$ \cite{cal_rev}. This form
implicitly assumes well localized magnetic moments leading to a
universal parametric relation between MAE ($K_{1}(T)$) and the
magnetization $M(T)$ \cite{cal_rev}.  The two-sublattice
modification of the QSPT proposed by Kuz'min \cite{kuz'min} has
been successfully applied to describe MAE of the localized 3d-4f
magnets.  Note, that all these theories predict that in the low
temperature region MAE scales as $K_{1}(T) \sim M^3(T)$ except the
very recent work by Skomski et. al.  where a  mean-field two
sub-lattice Hamiltonian for L1$_{0}$ CoPt led to a $K_{1}(T) \sim
M^2(T)$ dependence \cite{skomski_copt}.

Recent experimental results demonstrate that the uniaxial MAE of
epitaxial FePt films can be very accurately fitted to a $K_{1}(T) \sim
M^{2.1}(T)$ dependence in the low temperature range
\cite{thiele,okamoto}. This observation clearly demonstrates that the
contribution of the single-ion anisotropy (leading to the $M^{3}$
dependence) is practically missing. Thus, the Hamiltonian used in the
QSPT theory does not necessarily reflect all the essential features of
the magnetic interactions in L1$_{0}$ FePt and possibly also of the
other 3d-4d/5d ordered alloys.

In the following we present a model of magnetic interactions in FePt
which is constructed and parameterized on the basis of
first-principles calculations and is shown to be capable of explaining
on the quantitative level recent measurements of the $K_1 ~\sim
M^{\gamma}(T)$ dependence with non-integer exponent $\gamma =2.1$
\cite{thiele,okamoto}. Thus we propose a microscopic explanation of
this unusual behavior and test our microscopic model of magnetic
interactions.

The leading contribution to the anisotropic part of the spin
Hamiltonian is described as anisotropic exchange mediated by the
induced Pt atomic spin moments. The thermodynamic behavior of this
Hamiltonian is investigated within the mean-field approximation
(MFA) and in the classical limit using both Langevin dynamics and
Monte Carlo simulations. We find that proper treatment of the
magnetic interactions mediated by the induced Pt moment yield
$K_1(M(T))$ and $M(T)$ dependences in a good quantitative agreement
with experiment, including the value of $T_c$. Thus the proposed
atomic scale model describes correctly the most important static
magnetic properties and thus opens the way for modeling even more
complex dynamic switching properties \cite{our:sw}.

\section{Effective spin Hamiltonian model based on first-principles CLSDA calculations }
\begin{figure}[ht]
  \centering \includegraphics[bb= 50 230 600 770,width=8.2cm]{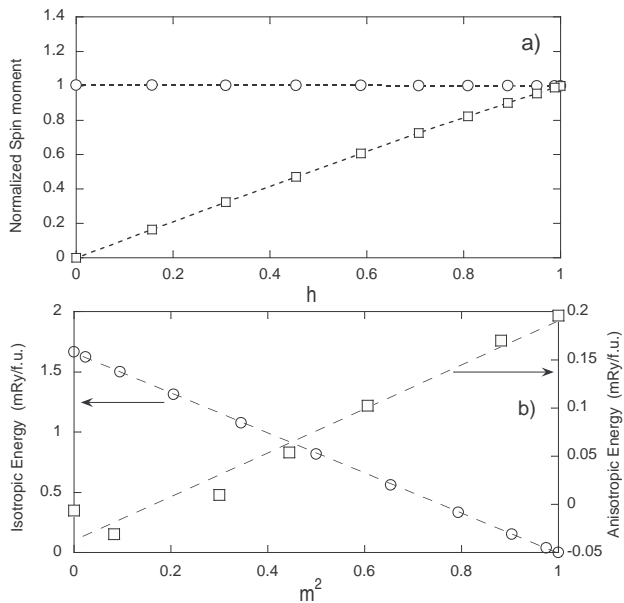}
  \caption{ Results of the constrained self-consistent LSDA
   calculations for ordered $L1_{0}$ FePt: a) Fe (circles) and Pt
   (squares) spin moments normalized by their values in the FM state
   as a function of normalized exchange field $h$; b) isotropic
   (squares) and anisotropic (circles) part of the total energy as a
   function $[\it{m}_{\nu}]^{2}$.  Dashed lines used for the linear
   fit.}
  \label{fig:noncol}
\end{figure}

Our analysis begins with an investigation of the isotropic part of the
spin Hamiltonian. We start with the constrained
local-spin-density-approximation (CLSDA) calculations
\cite{oleg_prb_92} for a non-collinear arrangement of Fe and Pt atomic
spin moments as summarized in Fig. \ref{fig:noncol}.  The CLSDA method
\cite{deder} allows to reduce the many electron problem to a
minimization of the Hohenberg-Kohn energy functional
$E_{HK}(\rho(\vec{r}), \vec{\sigma}(\vec{r}))$ of charge,
$\rho(\vec{r})$, and spin density, $\vec{\sigma}(\vec{r})$, with an
additional constraint term which in the case of a non-collinear
magnetic configuration leads to a CLSDA functional
$E_{CLSDA}(\rho(\vec{r}), \vec{\sigma}(\vec{r}), \vec{h}^{\perp}_{i})$
with an additional Lagrange multiplier ${\vec{h}}^{\perp}_{i}$ having
the meaning of an internal magnetic field \cite{oleg_prb_92}. This
magnetic field is determined self-consistently according to the
condition of the desired orientation of the atomic moment
${\vec{m}_{i}}$ at the site $i$.  The effect of thermal fluctuations
on the electronic and spin sub-systems can be conveniently modeled
with the spin-spiral (SS) configurations representing various degrees
of short range order which is found in 3d magnets for temperatures
well above $T_c$ \cite{oleg:heine:90}. The values of the Fe and Pt
moments for these SS configurations are calculated in the local
coordinate system, associated with the orientation of the local
quantization axis at site $i$ as $ M_{i}=\int_{- \infty}^{\it{e_{F}}}
\big[ n^{up}_{i}(\varepsilon) - n^{dn}_{i}(\varepsilon) \big]
d\varepsilon$, where the local density of electronic states is a
diagonal matrix over the spin indices $n_{i}^{\sigma, \sigma}$
\cite{sandra,oleg1}. To summarize the most important results obtained
for various SS configurations we introduce the convenient variable
$h_{\nu}=H_{\nu}/H_{\nu}^{0}$ where $H_{\nu}$ is the exchange field at
site $\nu$ of the Pt sublattice normalized by its value in the FM
state $H_{\nu}^{0}$.

In Fig.\ref{fig:noncol} we present the spin moments and total energies
of the non-collinear magnetic configurations calculated
self-consistently within the CLSDA, using a generalization of the
electronic structure method to treat non-collinear magnetic order
\cite{oleg_prb_92}. In Fig.\ref{fig:noncol}a we present normalized
values of the Fe and Pt spin moments calculated as a function of $h$.
The dependence shown in Fig. \ref{fig:noncol}a clearly demonstrates a
dramatic difference in the degree of localization for Fe and Pt spin
moments, respectively.  The Fe spin moment remains almost constant as
a function of $h$ (or angle $\theta$) indicating its relatively
localized nature in terms of the response to the thermal
fluctuations. On the other hand, the Pt moment varies linearly with
$h$. This result raises the important question how to develop an
adequate model to describe a system with mixed localized and
de-localized magnetic degrees of freedom.

The theory of magnetic interactions due to localized magnetic moments
is well established. Hence an effective spin Hamiltonian associated
with the localized (Fe moments) degrees of freedom can be constructed
in the form
\begin{eqnarray}
  \label{eq:mag-in-fe}
    H_{loc} = -\sum_{i \neq j}{J_{i j} {\bf S}_{i} \cdot {\bf S}_{j}}
    - \sum_{i} \it{k}_{Fe}^{(0)}
     [{\bf S}^{z}_{i}]^{2},
\end{eqnarray}
which relies on configuration independent effective exchange
interaction parameters $J_{i j}$ and an effective single-ion
anisotropy $\it{k}_{Fe}^{(0)}$. The ${\bf S}_{i}$ are used to denote
Fe sublattice spin moments which can be treated as unit vector in the
classical limit. The form of the spin Hamiltonian
Eq. \ref{eq:mag-in-fe} is well justified by our CLSDA results which
clearly indicate that the Stoner excitations associated with Fe
moments have much higher energy than those of the Pt.

On the other hand, the value of the induced Pt magnetic moment varies
between 0 and a maximum value for the FM state. The CLSDA total energy
calculations without s-o coupling (isotropic energy) presented in
Fig. \ref{fig:noncol}b allow to clarify how to deal with Pt magnetic
degrees of freedom.  Indeed, as the scalar-relativistic calculations
show, the total energy associated with these delocalized degrees
of freedom $E^{iso}_{deloc}$ follows very closely the relation which
can be derived from the Stoner-model expression for the total energy
\cite{moruzzi}
\begin{eqnarray}
  \label{eq:stoner}
  E_{deloc}^{iso} =
  \int_{- \infty}^{\it{e_{F}}} d\varepsilon
      [n^{up}(\varepsilon)+n^{dn}(\varepsilon)] \; \varepsilon
      -\frac{1}{2} I M_{pt}^2 \approx \tilde{I} [\bf{\it{m}}_{\nu}]^2
\end{eqnarray}
where $I$ is the intra-atomic exchange interaction parameter,
$M^{0}_{\nu}$ is the Pt magnetic moment in the FM state and $n^{up},
n^{dn}$ are spin resolved densities of states forming delocalized Pt
moments denoted as $M_\nu$ with $\tilde{I}_{\nu} = 1/2 I_{\nu}
[M^{0}_{\nu}]^{2}$ and $\it{m}_{\nu} = M_{\nu} / M^{0}_{\nu} $.
The dependence shown in Fig.~\ref{fig:noncol}a also demonstrates
that both the value and orientation of the Pt moment are entirely
due to the exchange field of the surrounding Fe moments, following
very closely the relation $\bf{M}_{\nu} = \chi_{\nu}
\bf{H}_{\nu}$, where $\chi_{\nu}$ is the local Pt susceptibility
constant. As follows from the linear dependence in
Fig.~\ref{fig:noncol}(a,b), $\it{m}_{\nu}$ can be very accurately
described by the relation,
\begin{eqnarray}
  \label{eq:m_pt}
   {\bf m}_{\nu}  = \frac{\chi_{\nu} } {M^{0}_{\nu}} \sum_{i} J_{i\nu} {\bf S}_{i}, 
\end{eqnarray}
where the Pt sublattice $\chi_{\nu}$ is found to have a weak
magnetic configuration dependence; the $J_{i\nu}$ are the effective
exchange parameters defined as the CLSDA total energy variation
$\delta E_{CLSDA}/\delta {\bf S}_i \delta {\bf m}_{\nu}$ in the FM
state, where ${\bf S}_{i}$ is the Fe and ${\bf m}_{\nu}$ the Pt
sub-lattice moment.
 The temperature
dependence of $\chi_{\nu}$ arising from the Fermi distribution
smearing is weak and will be neglected in the following
statistical simulations.

The anisotropic part of the effective spin Hamiltonian is calculated
within the CLSDA, with s-o interactions included self-consistently,
and is presented in Fig. \ref{fig:noncol}b.  We find that it can be be
very accurately approximated by a quadratic dependence on the
$m_{\nu}$ parameter. We should emphasize that owing to the delocalized
nature of 5d/4d elements, in the general case this dependence cannot
be guessed prior to the rigorous calculations.  However, our result
for FePt allows us to identify the form of the spin Hamiltonian
associated with delocalized magnetic degrees of freedom, $H_{deloc}=
H_{deloc}^{iso}+H_{deloc}^{anis}$.
\begin{eqnarray}
  \label{eq:h_pt}
     H_{deloc}  =  - \sum_{\nu} \tilde{I} {\bf m}_{\nu}^2  -
     \sum_{\nu} k^{(0)}_{Pt} ({\bf m}^z_{\nu})^2.
\end{eqnarray}
The magnetic energy is partitioned into localized and delocalized
contributions using the CLSDA approach allowing for a unified
description of the electronic degrees of freedom within the
one-electron approximation. In particular, the Fe $\it{k}^{(0)}_{Fe}$
and the Pt single-ion $\it{k}^{(0)}_{Pt}$ contributions can be
distinguished. In agreement with a previous study \cite{igor} we find
that the Fe contribution is negative while Pt gives rise to a large
easy axis contribution.  The Fe and Pt contributions to MAE have been
calculated within the LSDA and then corrected according to the
previous LSDA+U calculations\cite{mae_our}.  We find
${\it k}^{(0)}_{Pt}= 1.427$~meV and ${\it k}^{(0)}_{Fe}= - 0.097$ ~meV
which corresponds to the macroscopic uniaxial anisotropy constant $K_1
(T=0) = 7.7 ~ 10^7$ erg/cc.

Finally, with Eqs. (\ref{eq:mag-in-fe},\ref{eq:m_pt},\ref{eq:h_pt}) we
can introduce an effective spin Hamiltonian reflecting all the above
features revealed by our first-principles calculations.  It is
constructed as $H=H_{loc.}+H_{deloc}$ and can be reduced to the
convenient form
\begin{eqnarray}
  \label{eq:ham_all}
  H = - \sum_{i \ne j} \tilde{J}_{ij} {\bf S}_i \cdot {\bf S}_j
    - \sum_i d^{(0)}_i ({\bf S}^z_i)^2 - \sum_{i \ne j} d^{(2)}_{ij}
    {\bf S}^z_i {\bf S}^z_j.
\end{eqnarray}
We note that the spin Hamiltonian is now expressed in terms of the
Fe degrees of freedom, with effective exchange interaction
parameters $\tilde{J}_{ij} = J_{ij} + \tilde{I} (\frac{\chi_{\nu}
} {M^{0}_{\nu}})^{2} \sum_{\nu} J_{i \nu} J_{j \nu}$ and an
effective single -ion anisotropy,
\begin{eqnarray}
  \label{eq:eff_1ion}
 d^{(0)}_i = \it{k}_{Fe}^{(0)} + {\it{k}}_{Pt}^{(0)}
  (\frac{\chi_{\nu}} {M^{0}_{\nu}})^{2} \sum_{\nu} J_{i \nu}^2,
\end{eqnarray}
and a two-ion anisotropy contributions
\begin{eqnarray}
  \label{eq:eff_twoion}
d^{(2)}_{ij} = {\it{k}}_{Pt}^{(0)}
(\frac{\chi_{\nu}}{M^{0}_{\nu}})^{2} \sum_{\nu} J_{i\nu} J_{j
\nu}.
\end{eqnarray}
As can be seen from these expressions, the Pt induced spin moments
result in additional isotropic and anisotropic contributions, both
depending on the effective exchange interaction parameters $J_{i \nu}$
defined in the ferromagnetic ground state. We find that $J_{i \nu}$
are relatively strong and positive, resulting in isotropic and
anisotropic exchange interactions both stabilizing the ferromagnetic
order in the [001] direction (see \cite{note1}).  Unlike the $J_{i
\nu}$, the effective exchange interaction parameters between Fe
moments $J_{ij}$ appear to be sensitive to the lattice spacing and may
change sign from positive to negative as a function of the chemical
ordering \cite{brownPRB03}. Importantly, the additional anisotropic
contribution (Eqs.\ref{eq:eff_twoion}) which controls $K_{1}(M(T))$
does not depend on $J_{ij}$.

In order to assess the relative magnitudes of the single- and
two-ion terms (Eq. \ref{eq:ham_all}), consider for clarity the
nearest neighbor ($NN$) interaction only with N being the number of
$NN$'s. Then the  magnetic anisotropy free energy (F$_{anis}$(T))
within the first-order thermodynamic theory \cite{cal_rev}
(justified by $\langle H_{anis} \rangle /\langle H_{iso} \rangle
\approx 0.1$, see Fig. \ref{fig:noncol}b ) have an effective single
and two-ion contributions with latter   involving a sum over nearest
neighbors
\begin{eqnarray}
  \label{eq:fa}
F_{anis}(T) \approx \langle H_{anis}\rangle _{T} = d^{(0)}_i {\it
f}^{1}(T) + (N-1) d^{(2)}_{ij} {\it f}^{2}(T) ,
\end{eqnarray}
where
  ${\it f}^{1}(T) = \langle S_{i}S_{i}\rangle _{T}$ and
  ${\it f}^{2}(T) = \langle S_{i}S_{j} \rangle_{T}$ are the
  single-site and pair correlation functions,
  $d^{(0)}_i \approx  \it{k}_{Fe}^{(0)} + {\it{k}}_{Pt}^{(0)}/N$
  and
$d^{(2)}_i \approx  {\it{k}}_{Pt}^{(0)}/N$ within the $NN$
approximation. Given the small magnitude of $k^{(0)}_{Fe}$, the
ratio between the single and two-ion contributions $(N-1)
d^{(2)}/d^{(0)}\approx (N-1)$. Then, given the dominance of the
two-ion contribution, and since within the $MFA$ type approximation
${\it f}^{2}(T) \sim M^{2}(T)$, one can arrive at a qualitative
explanation of the observed MAE temperature dependence. Clearly, for
more accurate evaluation of the ratio between two and single-ion
contributions the distance dependence of the $J_{i \nu}$ has to be
taken into account. In the following we present calculations beyond
$MFA$ and $NN$ approximations to provide a quantitative analysis of
the proposed model in terms of its ability to explain the
non-integer exponent of the $K_1 \sim M^{2.1}$ dependence.

\begin{figure}[ht]
  \centering
  \includegraphics[width=7cm,clip]{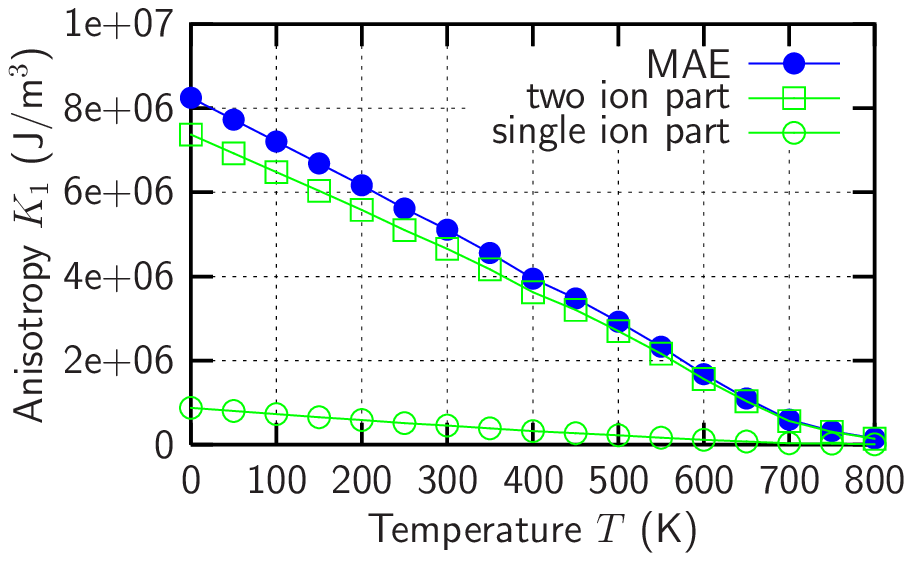}
  \includegraphics[width=7cm,clip]{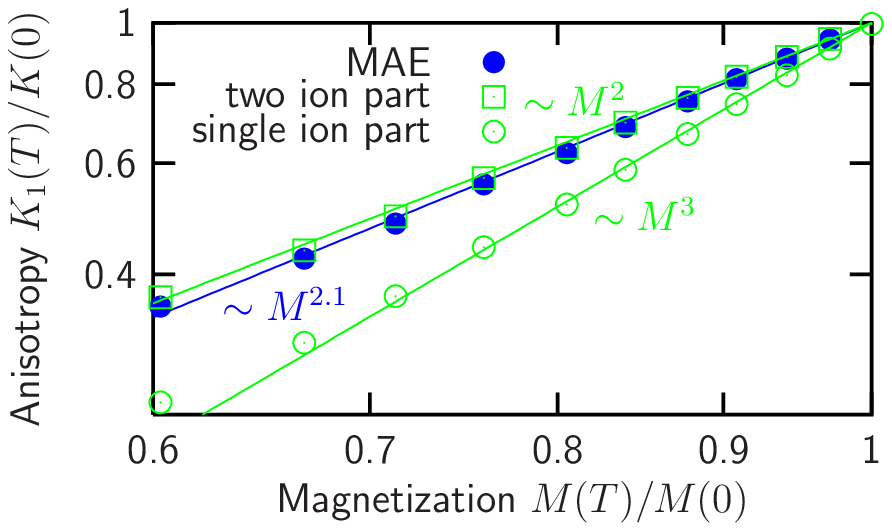} \\
  \includegraphics[width=7cm,clip]{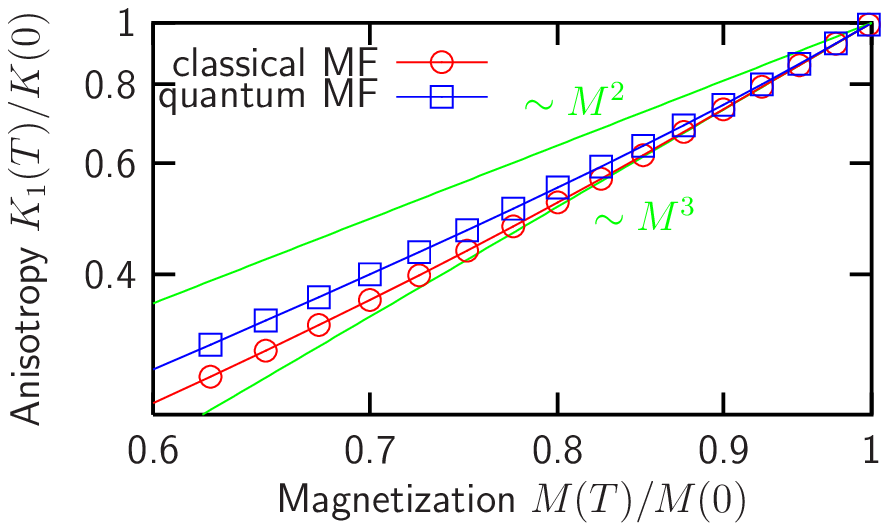}
  \includegraphics[width=7cm,clip]{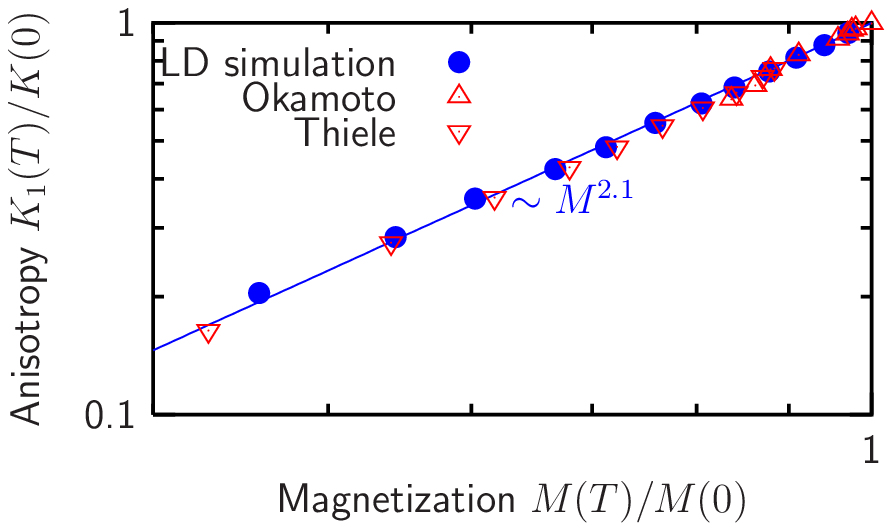}

  \caption {(a) $K_{1}(T)$ dependence using LD simulations with the
  effective spin Hamiltonian and its single and two-ion contributions;
  (b-d) log-log plots for $K_{1}(T)/K_{1}(0)$ vs. reduced
  magnetization $M(T)$: (b) using LD simulation within the classical
  approximation; c) calculated within the MFA QSPT for quantum $j =
  3/2$ and classical $j = \infty$, $M^2(T)$ and $M^3(T)$ dependence
  are presented for comparison; (d) comparision of $K_1(T)$
  calculated with LD and experimental data of Thiele et
  al. \cite{thiele} and Okamoto et al. \cite{okamoto}.  The solid
  lines gives a power law with exponent 2.1.}  \label{fig:KvsM}
\end{figure}

\section{Thermodynamic properties}
In the following, we use Langevin dynamics simulations described in
\cite{nowakARCP00} to investigate properties of the effective spin
Hamiltonian Eq.\ \ref{eq:ham_all} within the classical approximation.
The exchange interactions are long-ranged and are taken into account
for a distance of up to 5 atomic unit cells via
fast-Fourier-transformation, within the same calculation as the
dipolar interaction. We simulate spherical nanoparticles with open
boundary conditions and sizes up to 14464 moments, corresponding to
diameters up to 9.2nm.
The anisotropy constant defined as the free energy difference
between magnetization oriented parallel
or perpendicular
to the easy axis which according to the first order perturbation
theory \cite{cal_rev}  is given by the internal anisotropic energy
difference
$K_1(T) = E_a(T, \vec{B} = B \vec{e}_{\perp}) - E_a(T, \vec{B}
  = B \vec{e}_{||})$ for the external field  $\vec{B}$ .
Then, according to Eq.  \ref{eq:ham_all} single-ion and two-ion
contributions can be distinguished as summarized in Fig.\
\ref{fig:KvsM}a. One can see that the two-ion term is the dominant
contribution which is nearly nine times larger. Furthermore the data
indicate a Curie temperature close to the experimental value of 750K
\cite{okamoto,thiele}.
In Fig.\ \ref{fig:KvsM}b we present the calculated $K(M)$ dependence
along with its single and two-ion contributions.  Within these
calculations, which we stress go beyond MF classical approximation,
the two-ion term scales as $M^{2}(T)$ in a wide $T$ range, while the
single-ion term follows $M^{3}(T)$ scaling only at low temperatures.
Since $K_1(T)$ has both contributions, the expression for the low
$T$ expansion
\begin{equation}
  K_1(T)/K_1(0) \approx \alpha ~ M^2 + (1-\alpha) ~ M^3 \sim M^{3-\alpha},
\end{equation}
contains the $\alpha$ coefficient originating from the normalized
two-ion contribution and the second term originating from the
single-ion contribution.  The parameters $d^{(0)}_i$ and
$d^{(2)}_{ij}$ following from our first-principles calculations allow
us to evaluate finally the exponent of $3-\alpha = 2.09$.

Before proceeding to a comparison with experiments we examine the
range of validity of our classical statistical approximation.  In
Fig.\ \ref{fig:KvsM}c we present results of the MFA QSPT calculations
of the $K_1(M)$ dependence and corresponding two and single-ion
contributions.
Both, the classical and the quantum $K_1(M)$ dependence
are identical for the two-ion term in the whole range of temperatures
and for the single-ion term in the low temperature range. Considering
that the single-ion contribution is dominant, we can compare our
Langevin dynamics calculations with available experiment as shown in
Fig.\ \ref{fig:KvsM}d.  As one can see, our spin Hamiltonian with
ab-initio parameterization agrees very well in a wide range of
temperatures, especially given that the low temperature measurements by
Okamoto, et al. also yielded an exponent of 2.1 \cite{okamoto}.

\section{Summary and Conclusions}
To summarize, we propose an atomic-scale model of magnetic
interactions in ordered L1$_{0}$ FePt with an effective spin
Hamiltonian constructed and parameterized on the basis of
first-principles calculations. The proposed model is investigated
analytically and using statistical simulations. We find that the
model describes on the quantitative level the experimentally
observed anomaly in the temperature dependence of the magnetic
anisotropy energy.  We demonstrate that this observed, anomalous
temperature dependence ($K_1 \sim M^{2.1}(T)$) is due to the
delocalized induced Pt moments, leading to an exchange mediated
two-ion anisotropy which dominates the usually expected $M^3$
contribution of the single-ion anisotropy. We believe that this
mechanism is common for various 3d-5d/4d ordered alloys having 5d/4d
nominally non-magnetic elements with large s-o coupling and Stoner
enhanced susceptibility.

\acknowledgments We thank R. Skomski, A. Shick, M. van Schilfgaarde,
T. Schulthess, R. Sabirianov, C. Platt and D. Weller for useful and
stimulating discussions.

\bibliographystyle{draft} \bibliography{biblio}

\end{document}